\documentclass[journal=nalefd, manuscript=article]{achemso}

\usepackage{amsmath}
\usepackage{siunitx}
\usepackage{bbold}
\usepackage{mathtools}

\DeclareSIUnit\dBm{dBm}

\author{A.~Iorio}
\email{andrea.iorio@sns.it}
\affiliation{NEST, Istituto Nanoscienze-CNR and Scuola Normale Superiore, I-56127 Pisa, Italy}
\author{A.~Crippa}
\affiliation{NEST, Istituto Nanoscienze-CNR and Scuola Normale Superiore, I-56127 Pisa, Italy}
\author{B.~Turini}
\affiliation{NEST, Istituto Nanoscienze-CNR and Scuola Normale Superiore, I-56127 Pisa, Italy}
\altaffiliation{Present Address: ICFO - Institut De Ciencies Fotoniques, The Barcelona Institute of Science and Technology, 08860 Castelldefels (Barcelona), Spain}
\author{S.~Salimian}
\affiliation{NEST, Istituto Nanoscienze-CNR and Scuola Normale Superiore, I-56127 Pisa, Italy}
\author{M.~Carrega}
\affiliation{CNR-SPIN, Via Dodecaneso 33, 16146 Genova, Italy}
\author{L.~Chirolli}
\affiliation{NEST, Istituto Nanoscienze-CNR and Scuola Normale Superiore, I-56127 Pisa, Italy}
\author{V.~Zannier}
\affiliation{NEST, Istituto Nanoscienze-CNR and Scuola Normale Superiore, I-56127 Pisa, Italy}
\author{L.~Sorba}
\affiliation{NEST, Istituto Nanoscienze-CNR and Scuola Normale Superiore, I-56127 Pisa, Italy}
\author{E.~Strambini}
\affiliation{NEST, Istituto Nanoscienze-CNR and Scuola Normale Superiore, I-56127 Pisa, Italy}
\author{F.~Giazotto}
\affiliation{NEST, Istituto Nanoscienze-CNR and Scuola Normale Superiore, I-56127 Pisa, Italy}
\author{S.~Heun}
\email{stefan.heun@nano.cnr.it}
\affiliation{NEST, Istituto Nanoscienze-CNR and Scuola Normale Superiore, I-56127 Pisa, Italy}

\title{Half-integer Shapiro steps in highly transmissive InSb nanoflag Josephson junctions}

\begin{document}

\begin{tocentry}
\includegraphics{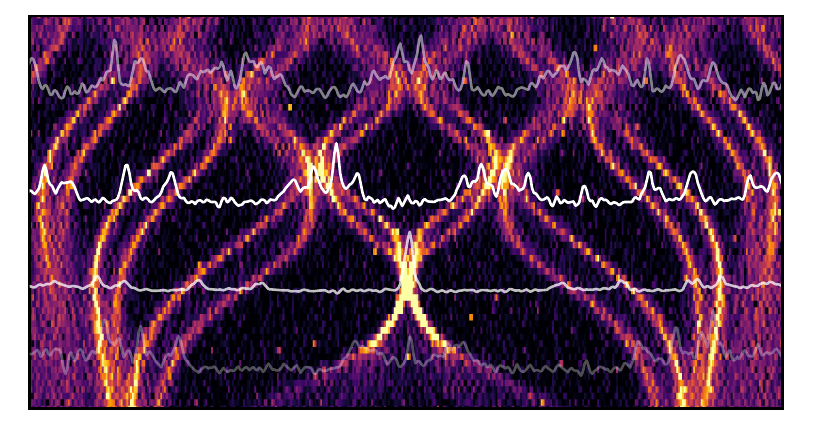}
\end{tocentry}

\begin{abstract}
We investigate a ballistic InSb nanoflag-based Josephson junction with Nb superconducting contacts. The high transparency of the superconductor-semiconductor interfaces enables the exploration of quantum transport with parallel short and long conducting channels. Under microwave irradiation, we observe half-integer Shapiro steps that are robust to temperature, suggesting their possible non-equilibrium origin. Our results demonstrate the potential of ballistic InSb nanoflags Josephson junctions as a valuable platform for understanding the physics of hybrid devices and investigating their non-equilibrium dynamics.
\end{abstract}

The advancing quantum technologies have made the investigation of low-dimensional hybrid superconducting nanostructures a major area of research in recent years. When a normal conductor is coupled to a superconductor, the superconducting correlations can penetrate into the non-superconducting region through the proximity effect~\cite{kulik_1969, pannetier_2000}. As a result, the hybrid system can exhibit unique properties derived from both the normal and superconducting components, offering exciting possibilities for novel functionalities. This phenomenon has been investigated in various solid-state platforms, including semiconductors~\cite{jarillo-herrero_2006, calado_2015}, two-dimensional electron systems~\cite{lehnert_1999, giazotto_2004, ke_2019}, magnetic and ferroelectric materials~\cite{sellier_2004, frolov_2006} and topological insulators~\cite{veldhorst_2012, hart_2014, pribiag_2015}. In this context, Indium Antimonide (InSb) is a particularly promising semiconductor, known for its high electron mobility, narrow bandgap, strong Rashba spin-orbit coupling, and large $g^*$-factor~\cite{qu_2016, ke_2019, mayer_2020, moehle_2021, lei_2021}. Due to the challenges of growing InSb quantum wells on insulating substrates, free-standing InSb nanoflags have emerged as a highly flexible platform, as they can be grown without defects on lattice-mismatched substrates~\cite{chen_2021, delamata_2016, devries_2019, pan_2016, zhi_2019j, zhi_2019i, verma_2020, kang_2019, xue_2019, chen_2021b, gazibegovic_2019, verma_2021}. They have also been referred to as nanoflakes, nanosheets, or nanosails in the literature~\cite{delamata_2016, pan_2016, rossi_, devries_2019}. Recently, InSb nanoflags have been used to realize proximity-induced superconductor-normal metal-superconductor (SNS) Josephson junctions, which exhibit ballistic and gate-tunable supercurrents~\cite{zhi_2019j, salimian_2021, turini_2022}, clear subharmonic gap structures~\cite{zhi_2019i, salimian_2021}, and non-local and non-reciprocal supercurrent transport~\cite{turini_2022, devries_2019}. These developments highlight the potential of InSb nanoflags as a platform for exploring the complex dynamics between charge, spin, and superconducting correlations, including topological superconductivity~\cite{fornieri_2019, prada_2020}, gate-tunable hybrid superconducting qubits~\cite{larsen_2015, casparis_2018, hays_2021}, and non-equilibrium quasiparticle dynamics~\cite{basset_2019, catelani_2019, hays_2021}. 

In this work, we present a thorough investigation of highly transmissive ballistic Josephson junctions on InSb nanoflags made with niobium (Nb) contacts. Compared to previous works~\cite{salimian_2021, turini_2022}, our device has a higher junction transparency, which enables the investigation of unexplored transport regimes. Our findings reveal the coexistence of parallel short and long conducting channels, as confirmed by the temperature dependence of the critical current and magnetoresistance. Under microwave irradiation, we observe Shapiro steps at half-integer values of the canonical voltage $hf/2e$, which exhibit a non-monotonic evolution with temperature. The observation suggests that a non-equilibrium state is formed in the junction due to the microwave drive.

\begin{figure*}
 \includegraphics[width=\textwidth]{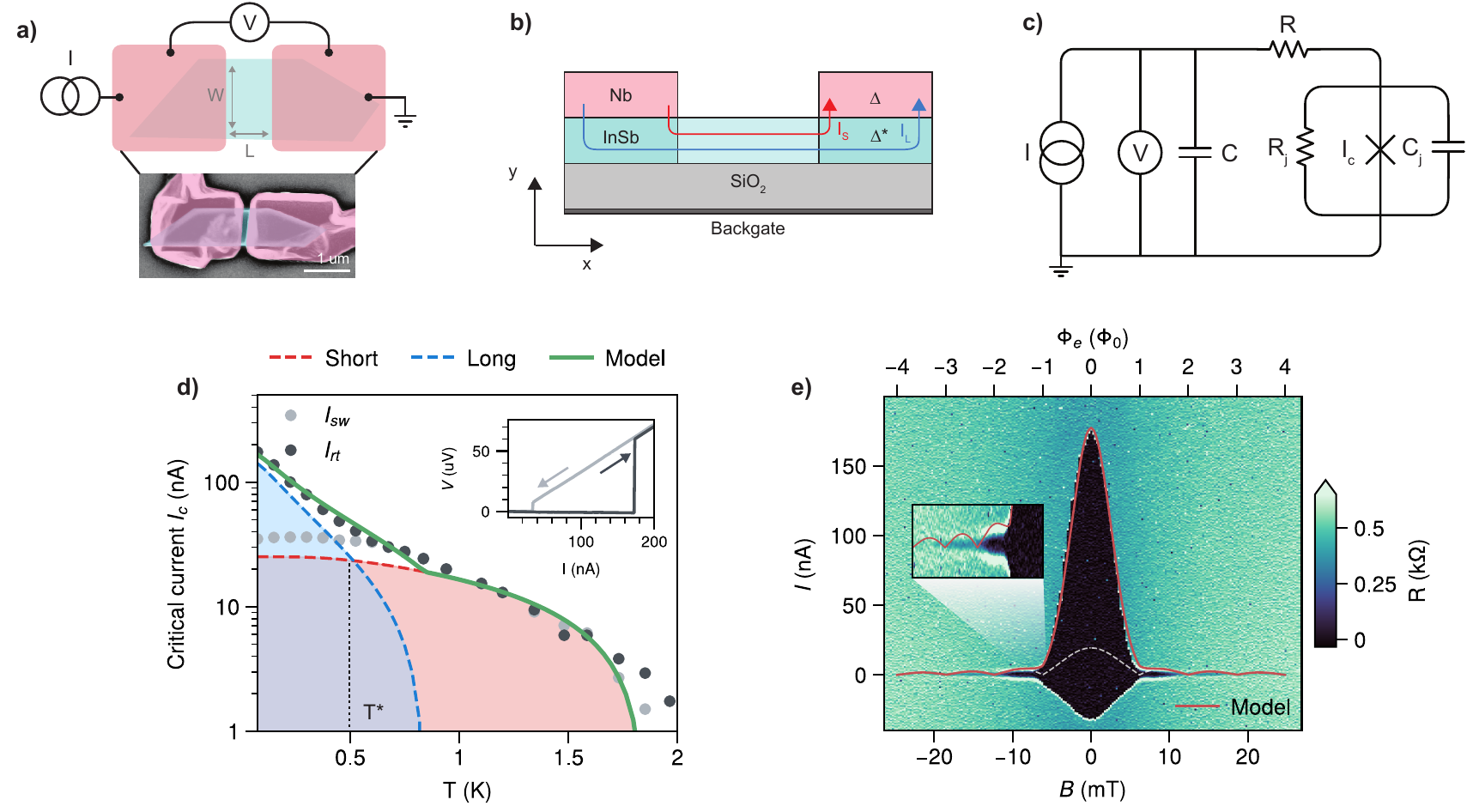}
\caption{(a) Upper part: sketch of the sample with the relevant dimensions and a simplified measurement setup. The junction length is $L=\SI{80}{\nm}$ and the width is $W=\SI{650}{\nm}$. Lower part: scanning electron micrograph of the SNS junction. The InSb nanoflag has a trapezoidal shape, and the Nb contacts are patterned on top of it. (b) Schematic cross-section of the device, where the superconducting Nb contacts with gap $\Delta$ proximitize an induced gap $\Delta^*$ in the InSb layer.~\cite{kjaergaard_2017} The red and blue lines represent, respectively, the short and long conducting channels that are discussed in (d). (c) Extended RCSJ model, with the Josephson junction of critical current $I_c$ in parallel with a shunt resistance $R_j$ and capacitance $C_j$. Additional shunt capacitance $C$ and resistor $R$ take into account the dissipative environment around the junction. (d) Temperature dependence of the switching current $I_{sw}$ (black dots) and the retrapping current $I_{rt}$ (grey dots). The blue and red areas indicate the contributions of the long and short conducting channels, respectively, as estimated from the corresponding models. The green line represents the sum of the two contributions. Inset: forward (black line) and backward (grey line) current sweeps used to extract the switching $I_{sw}$ and retrapping $I_{rt}$ currents, respectively. (e) Differential resistance $R=dV/dI$ plotted as a function of the bias current $I$ and the out-of-plane magnetic field $B$. The red curve takes into account both the long and short model contributions. The grey line shows the Fraunhofer pattern resulting from the short channel only. The inset provides a zoomed-in view of lobes in the low-bias region.} 
\label{fig1}
\end{figure*}

The device depicted in Figure~\ref{fig1}a and \ref{fig1}b consists of a planar SNS junction made of an InSb nanoflag with two Nb contacts. Previous studies have revealed that these InSb nanoflags are defect–free and exhibit a zincblende structure with high mobility (up to \SI{29500}{\cm\square\per\V\per\second}) and a mean free path $l_e \simeq 500$~nm at \SI{4.2}{\K}~\cite{verma_2021}. The Fermi wavelength $\lambda_F \simeq 30$~nm for a carrier concentration of $n_{s} \simeq 8.5 \times 10^{11}$~cm$^{-2}$ is comparable to the thickness of the nanoflags ($\simeq 100$~nm), resulting in a strong quasi two-dimensional character. Measurements were performed using a standard four-wire technique at the base temperature of $T = 75$~mK of a dilution refrigerator. A highly doped Si backgate allows for control of the carrier density of the InSb and was set to $V_{G} = \SI{40}{\V}$ for the results shown in the following. Microwave signals are applied via an open-ended attenuated coaxial cable placed $\sim \SI{1}{\cm}$ away from the chip surface. The junction dynamics is modeled using an extended resistively and capacitively shunted junction (RCSJ) model, which takes into account the dissipative environment surrounding the junction, as depicted in Fig.~\ref{fig1}c~\cite{russer_1972, jarillo-herrero_2006, larson_2020}. Further information on materials, fabrication, and measurement techniques can be found in the Supporting Information.

We first characterize the device in the absence of microwave irradiation. A typical back and forth sweep $V(I)$ is presented in the inset of Figure~\ref{fig1}d, in which a current bias $I$ is applied and the resulting voltage drop $V$ across the junction is measured. The $V(I)$ characteristics shows a considerable hysteresis with a switching current $I_{sw} \simeq \SI{170}{\nA}$ and a retrapping current $I_{rt} \simeq \SI{30}{\nA}$. The hysteresis in planar SNS junctions is commonly due to electronic heating in the normal region~\cite{courtois_2008}, with a finite junction capacitance $C_j$ potentially contributing~\footnote{In our device, the geometric junction capacitance is estimated $\sim$ aF, which may not cause a noticeable hysteresis, but intrinsic capacitance effects cannot be ruled out~\cite{antonenko_2015, massarotti_2018, fischer_2022}.}.

Figure~\ref{fig1}d shows the temperature dependence of the switching and retrapping currents on a semi-log scale. We can distinguish two distinct regions in the data. For temperatures $T > T^*$, with $T^* \sim \SI{500}{\milli\K}$ (see Figure~\ref{fig1}d), the switching current follows the predictions of a short junction model (shown as a red shaded area). However, for $T \leq T^*$, we see a deviation from the short junction behavior, and the switching current follows an exponential increase with decreasing $T$, which is characteristic of long junctions (blue shaded area). The data can be well-reproduced over the entire temperature range using a simple model that considers the transport predominantly determined by two conducting channels, long and short, as illustrated in Figure~\ref{fig1}b and demonstrated by the green line in Figure~\ref{fig1}d.

In the short junction limit $L \ll \xi_N$ (where $\xi_N = \hbar v_F/\Delta \simeq \SI{720}{\nm}$ is the coherence length, with $v_F \simeq 1.5 \times 10^6$~m/s \cite{verma_2021} and $\Delta \simeq \SI{1.35}{\meV}$), the supercurrent flows directly through the InSb region between the Nb contacts separated by $L = \SI{80}{\nm}$. For simplicity, we assume that all modes in the junction have equal effective transmission $\tau$, which can be described in the ballistic limit ($L \ll l_e$) by~\cite{beenakker_1991}:
\begin{equation}
 I_{S} (T) = \max_{\varphi} \frac{\overline{N} e\Delta^{*^2}(T)}{2\hbar}\frac{ \tau \sin \varphi}{E_A(\varphi, T)} \tanh{\frac{E_A(\varphi, T)}{2 k_B T}},
\label{eq1}
\end{equation}
with $\overline{N}$ the number of effective modes, $E_A(\varphi, T) = \Delta^* (T) \sqrt{1-\tau \sin^2(\varphi/2)}$ the Andreev bound state (ABS) energy of the mode, $\Delta^*(T) = \Delta^*(0) \tanh\left(1.74\sqrt{T_c/T-1}\right)
$ the tem\-per\-a\-ture-dependent induced energy gap,\footnote{We are assuming for simplicity a BCS gap.}\cite{muehlschlegel_1959} and $\varphi$ the macroscopic phase-difference across the junction. The best fit with the short junction model yields the red dashed line in Figure~\ref{fig1}d, with a single mode $\overline{N} = 1$, $\tau=0.93$ and $T_c = \SI{1.85}{\K}$, and a value of critical current $I_S = \SI{25}{\nA}$ at $T=\SI{75}{\milli\K}$. The observed lower values of currents are consistent with the transport mechanism illustrated in Fig.~\ref{fig1}b, where the supercurrent flows between the two proximized InSb regions with an induced gap $\Delta^*$, rather than being dominated by the Nb gap $\Delta$.

The exponentially enhanced conduction at low temperatures is typical of long channel states (of length $d$). The conduction via these states holds in the long junction limit $d \gg \xi_N$, and reads~\cite{dubos_2001}:
\begin{equation}
 I_{L} (T) = \frac{E_{Th}}{R_Ne}a\left [1 - 1.3\exp\left(-\frac{aE_{Th}}{3.2k_BT}\right)\right],
\end{equation}
where $E_{Th} = \hbar v_F l_e /2d^2$ is the Thouless energy~\cite{pannetier_2000}, $R_N$ is the junction resistance and $a= 3$.~\footnote{The constant $a$ is dependent on the ratio $E_{Th}/\Delta^*$.~\cite{dubos_2001}} The best fit of the long junction model is shown as the blue dashed line in Figure~\ref{fig1}d and yields $E_{Th} \sim \SI{20}{\micro\eV}$, corresponding to $d \sim \SI{3.5}{\um}$, close to the total length of the InSb nanoflag (\SI{3.35}{\um}), $R_N \sim \SI{400}{\ohm}$ and a critical current $I_{L} \sim \SI{140}{\nA}$ at $T = \SI{75}{\milli\K}$. 

Previous studies have documented similar results in highly transmissive ballistic SNS junctions with topological insulators or graphene~\cite{calado_2015, borzenets_2016, kayyalha_2019, schuffelgen_2019, stolyarov_2020, rosenbach_2021, schmitt_2022}, with the behavior being attributed to contributions from both surface and bulk states~\cite{schuffelgen_2019, rosenbach_2021}. One study linked the low-temperature enhancement to a low-energy Andreev bound state localized around the circumference of the junction~\cite{kayyalha_2019}. In our nanoflags, this could be consistent with electronic transport at the edges of the nanoflag due to band-bending, similarly to what has been reported by~\citet{devries_2019}. Compared to earlier works on InSb nanoflags that employed a Ti sticking layer between Nb and InSb~\cite{salimian_2021, turini_2022}, the increase of $I_{sw}$ at low temperature is consistent with the increased transparency achieved in this study through the direct deposition of bare Nb on the passivated surface of InSb, without the use of additional metallic layers. 

Magnetotransport measurements further confirm the coexistence and magnitude of the two current conducting channels in the junction, providing additional insight into the current density distribution across the channels. The differential resistance of the junction $R = dV/dI$ as a function of magnetic flux is presented in Figure~\ref{fig1}e. An unconventional Fraunhofer pattern, with a first lobe much more pronounced than the side lobes, is visible and well-described by the superposition of a conventional Fraunhofer pattern typical of short junctions, and a monotonic quasi-Gaussian decay, which is characteristic of long SNS junctions~\cite{barzykin_1999, cuevas_2007, angers_2008, chiodi_2012, stolyarov_2020, blom_2021}. The forward-biasing of the current results in a non-symmetrical supercurrent region for switching and retrapping currents (black area). The periodicity of the Fraunhofer pattern corresponds to one flux quantum inside the junction, taking into account a London penetration depth of $\lambda_L \simeq \SI{100}{\nm}$~\cite{gubin_2005} and a flux enhancement of a factor of $\Gamma_f \sim 1.8$ due to flux focusing within the planar geometry. The critical current values from short and long transport channels estimated in Figure~\ref{fig1}d are used here to model the magnetic interference patterns. The red line in Figure~\ref{fig1}e shows the combined contribution of both channels to the supercurrent $I (\Phi_{e}) = I_{S} (\Phi_{e}) + I_{L} (\Phi_{e})$, where $\Phi_e = \Gamma_f B(L+2\lambda_L)W$ is the applied magnetic flux on the uncovered junction area, with $W = \SI{650}{\nm}$ the junction width. The standard Fraunhofer pattern $I_S(\Phi_{e}) = I_{S}| \sin(\pi (\Phi_{e}/\Phi_0))/(\pi \Phi_{e}/\Phi_0)|$ expected for a wide-short junction, and a Gaussian decay $I_L (\Phi_{e}) = I_{L} \exp (-\sigma \Phi_{e}^2/\Phi_0^2)$, typical of a narrow-long junction, are accounted for in the calculation. We have included a possible different effective area of the long junction directly in the estimated value of $\sigma \sim 0.329$ while preserving the same flux dependence. 

Our conclusions are further supported by the temperature-dependent change in the magnetoresistance, which exhibits an exponential reduction of the Gaussian component and limited variation in the Fraunhofer lobes up to $T = \SI{800}{\milli\K}$ (refer to Figure~S3 of the Supporting Information). The lack of distinct oscillations in the magnetoresistance indicates that possible edge states are not interfering coherently with magnetic fields perpendicular to the nanoflag. However, the impact of flux screening, phase decoherence, and transport along various facets of the flag make it challenging to arrive at more definitive conclusions.

\begin{figure*}
 \includegraphics[width=\textwidth]{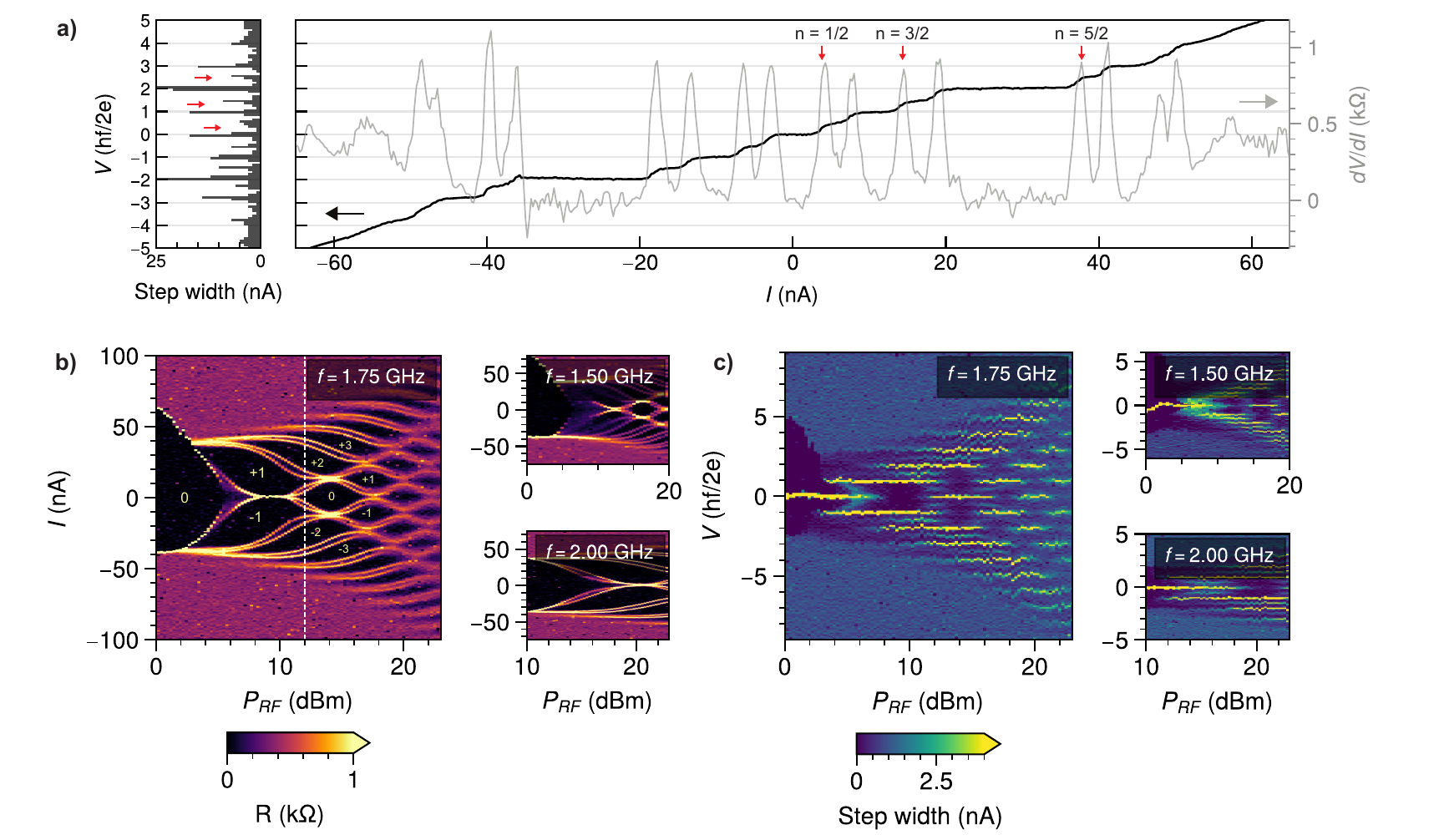}
\caption{(a) Sample $V(I)$ curve (black) in the presence of microwave irradiation at $P_{RF} = \SI{12}{\dBm}$ and frequency $f = \SI{1.75}{\GHz}$. The grey line represents the differential resistance $R=dV/dI$ with arrows highlighting half-integer Shapiro steps. The histogram on the left shows the distribution of the voltage data. The bin unit is equal to the current step size, such that the number of counts corresponds to the width of the voltage plateaus. The value of $P_{RF}$ refers to the one provided by the signal generator. (b) Full evolution of $R$ as a function of current bias $I$ and microwave power $P_{RF}$ at $f=\SI{1.75}{\GHz}$. Pairs of bright peaks indicate the presence of half-integer steps. The white dashed line corresponds to the data shown in (a). Label numbers refer to the corresponding step index $n$. The right side shows the colormaps for $f=\SI{1.50}{\GHz}$ and $f=\SI{2.00}{\GHz}$. (c) Histograms, as shown in (a), are displayed based on the data in (b) in a colorplot as a function of microwave power $P_{RF}$.}
 \label{fig2}
\end{figure*}

Having established the response of the junction at equilibrium, we will now examine how the system behaves when subjected to a microwave irradiation. In Figure~\ref{fig2}a, we present a sample $V(I)$ curve with a microwave tone at frequency $f = \SI{1.75}{\GHz}$ and applied power $P_{RF} = \SI{12}{\dBm}$. As it can be difficult to estimate the precise power delivered to the sample, we will only refer to the power provided by the signal generator in the following discussion. Quantized voltage steps of amplitude $n \times hf/2e$ appear in the $V(I)$ characteristic (black line), as a result of the phase-locking between the microwave frequency and the junction Josephson frequency~\cite{shapiro_1963}. In addition to integer steps occurring at $n = \pm 1, \pm 2, \pm 3, \ldots$, half-integer steps appear with $n = \pm 1/2, \pm 3/2, \pm 5/2, \ldots$~. The overlapping grey trace displays $dV/dI$ and shows peaks associated to both integer and half-integer steps, some of which are highlighted by red arrows. Steps with fractions different from multiples of $1/2$ are not observed. A histogram, resulting from the binning of the voltage data, is shown on the left and provides an immediate visual representation of the length of each step. The bin unit equals the current step size, such that the number of counts corresponds to the width of the voltage plateaus.

Figure~\ref{fig2}b shows a color plot of $R$ as a function of $I$ and $P_{RF}$. Sharp jumps in voltage appear as bright peaks in $R$, while voltage plateaus corresponds to dark regions. The pattern of bright peak pairs in the data provides stark evidence of fractional steps that occur over a wide range of power and frequencies, as demonstrated by the maps at $f = \SI{1.50}{\GHz}$ and $f = \SI{2.00}{\GHz}$. The region between the plateaus $\pm 1$ displays bistability at around $P_{RF} = \SI{10}{\dBm}$ of applied power, with sudden switching occurring between the two overlapping plateaus (see the Supporting Information for additional discussion). Figure~\ref{fig2}c better highlights the emergence of Shapiro steps by depicting the evolution of the histogram data as a function of microwave power for various frequencies. In the Supporting Information, we present additional measurements for different backgate voltages, magnetic field values, and temperatures. 

\begin{figure*}
 \includegraphics[width=\textwidth]{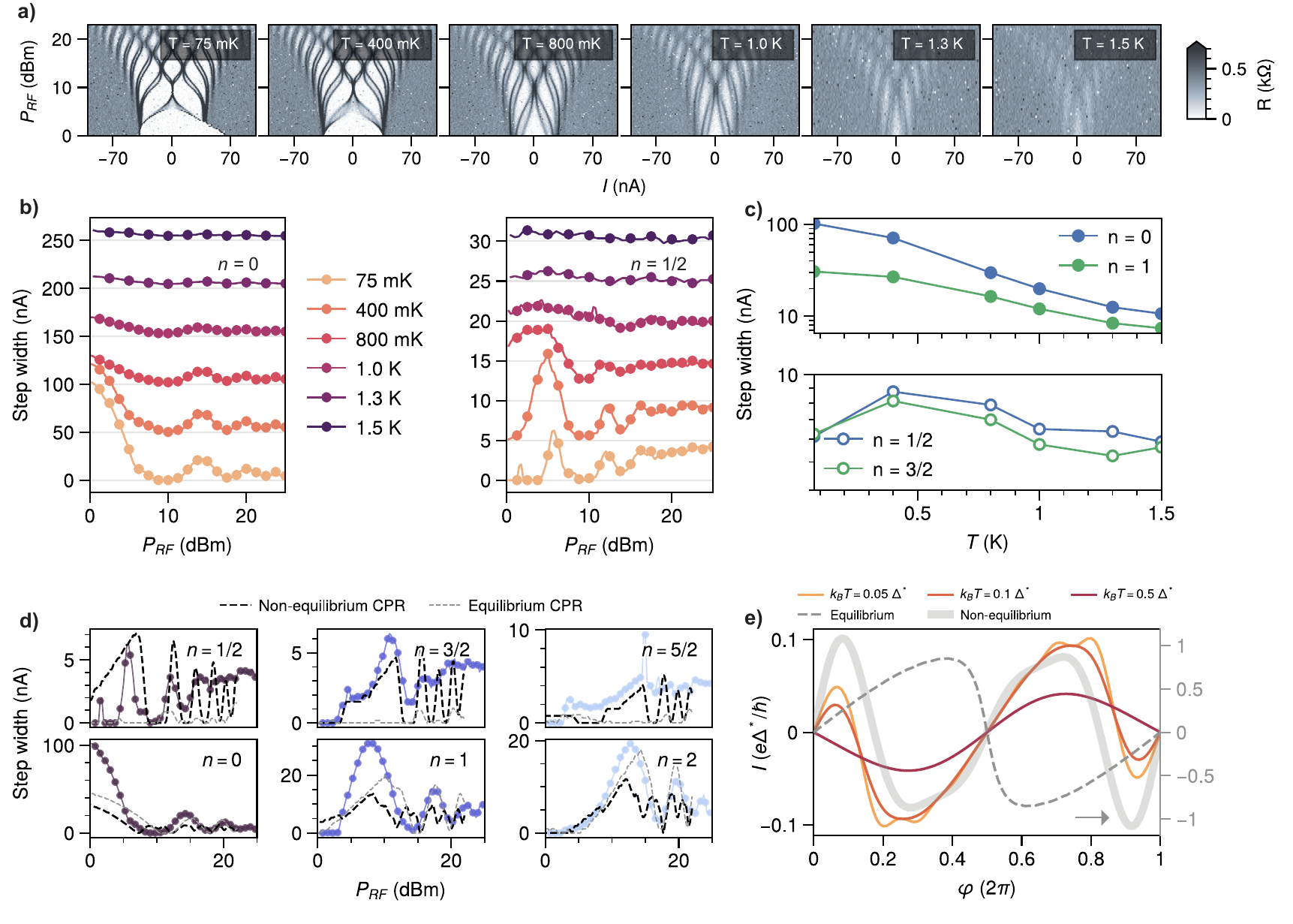}
\caption{(a) Shapiro maps at different temperatures $T =$ (\SI{75}{\milli\K}, \SI{400}{\milli\K}, \SI{800}{\milli\K}, \SI{1.0}{\K}, \SI{1.3}{\K}, \SI{1.5}{\K}) at $f = \SI{1.75}{\GHz}$. (b) Step width for $n=0$ (left) and $n=1/2$ (right) extracted from the temperature maps show in (a). As temperature increases, a monotonic decrease is observed for $n=0$, while a non-monotonic evolution with a maximum at \SI{400}{\milli\K} is observed for $n=1/2$. Traces are offset for clarity, with an offset of \SI{50}{\nA} (left) and \SI{5}{\nA} (right). (c) Step widths displayed on a semi-log scale as a function of $T$ for integers $n=0,1$ (top) and half-integers $n=1/2, 3/2$ (bottom). Microwave powers $P_{RF} = \SI{0}{\dBm}$ for $n=0$, $P_{RF} = \SI{8}{\dBm}$ for $n=1$, $P_{RF} = \SI{5.5}{\dBm}$ for $n=1/2$, and $P_{RF} = \SI{11}{\dBm}$ for $n=3/2$. (d) Step width of half-integer $n=1/2, 3/2, 5/2$ (top) and integer $n=0,1,2$ (bottom). Each trace is a horizontal slice of Figure~\ref{fig2}c. The dashed black lines represent the numerical simulations obtained from the extended RCSJ model using the non-equilibrium CPR shown as the thick grey line in (e). The dashed grey lines represent the simulations using the equilibrium CPR. (e) CPRs under microwave irradiation for $\tau = 0.98$ and driving $w = 1.3$ at different temperatures ($k_BT =$ 0.05, 0.1, and 0.5~$\Delta^*$) are depicted by light to dark orange lines. The dashed grey line represents the equilibrium CPR at $w=0$ and $k_BT = 0.05~\Delta^*$, for the same value of $\tau$. The thick grey line represents the effective non-equilibrium CPR used in panel (d).}
 \label{fig3}
\end{figure*}

We then study the behavior of the system by increasing the temperature. Figure~\ref{fig3}a displays Shapiro maps at temperatures ranging from $T = \SI{75}{\milli\K}$ ($\sim 0.02~\Delta^*/k_B$) to $T = \SI{1.5}{\K}$ ($\sim 0.5~\Delta^*/k_B$). The temperature rise leads to a decrease of the supercurrent and an increase in thermal fluctuations, resulting in rounded voltage plateaus. However, the half-integer steps remain stable from base temperature up to around \SI{1}{\K} ($\sim 0.3~\Delta^*/k_B$), where the current-phase relationship (CPR) given by the equilibrium supercurrent is expected to be mostly sinusoidal (as illustrated in Figure~S10). 

Figure~\ref{fig3}b shows the change in step width for the steps $n=0$ and $n=1/2$ extracted from Figure~\ref{fig3}a. The amplitude of the integer step decreases monotonically with increasing temperature, while the half-integer step shows a non-monotonic trend, with a maximum at $T \simeq \SI{400}{\milli\K} \sim 0.12~\Delta^*/k_B$. This is also demonstrated in Figure~\ref{fig3}c, where the step widths for $n=0,1$ and $n=1/2,3/2$ are plotted on a semi-log scale as a function of $T$. While the integer steps show an exponential decrease, the width of the half-integer steps first increases, then decreases, and eventually saturates due to the noise level at high temperatures. This remarkable evolution points to a non-equilibrium origin of the half-integer steps and is consistent with predictions and experimental observations that non-equilibrium supercurrents are less rapidly impacted by temperature compared to equilibrium supercurrents, which are suppressed exponentially~\cite{zaikin_1983, argaman_1999, lehnert_1999, dubos_2001a, fuechsle_2009, dassonneville_2018}.

In the Supporting Information, we provide data from an additional device with lower transparency. While the junction behaves similarly under microwave irradiation, the signatures of the half-integer steps are considerably weaker.

Despite their frequent occurrence, the origin of fractional steps in superconducting devices is not univocal. Measurements of fractional Shapiro step are commonly used to identify non-sinusoidal CPRs in highly transparent SNS junctions~\cite{ueda_2020}, or in junctions incorporating ferromagnetic layers~\cite{sellier_2004, pfeiffer_2008, frolov_2006, stoutimore_2018} or those exhibiting exotic superconducting states~\cite{shvetsov_2018, trimble_2021}. Geometric or intrinsic capacitance~\cite{eckern_1984, angers_2008, antonenko_2015, massarotti_2018, fischer_2022}, and circuit feedback~\cite{russer_1972, hamilton_1972, larson_2022} can also contribute to the appearance of fractional steps or hysteretic behavior. Sub-harmonic structures may also indicate a unique mode of a more complex circuit network, as seen in junction arrays~\cite{lee_1990, lee_1991, heinz_1997, valizadeh_2007, panghotra_2020, amet_2022} and superconducting quantum interference devices (SQUIDs)~\cite{vanneste_1988, early_1995, heinz_1997}. These manifestations are also visible even in the absence of multiple superconducting terminals, as in grain boundary or step-edge junctions~\cite{early_1993, terpstra_1995, ku_1995, yang_1994}, as a consequence of the complex evolution of multiple phase-locked states. 

The fractional steps reported in the previous examples, including both ballistic and diffusive SNS junctions, are ascribed to the equilibrium properties of the supercurrent and can be understood within a phenomenological extended resistively and capacitively shunted junction (RCSJ) model, which takes into account the dissipative environment surrounding the junction, as depicted in Fig.~\ref{fig1}c\cite{russer_1972, jarillo-herrero_2006, larson_2020}. In the phase-particle picture, neglecting capacitive effects, the phase evolves in a washboard potential that is tilted by the applied bias current and modulated by the time-dependent drive. Shapiro steps arise as time-dependent phase slips between the minima of the Josephson potential, and for a typical $\sin(\varphi)$ CPR, integer steps arise as $2\pi n$ phase slips. Within this picture, half-integer Shapiro steps require an energy-phase relation that displays a secondary minimum and arise when the second harmonic of the CPR is stronger than the first one.

However, a microwave drive can also significantly alter the supercurrent's steady-state behavior.~\cite{zaikin_1983, argaman_1999, kroemer_1999, lehnert_1999, biedermann_2001, baselmans_2002, cuevas_2002, jacobs_2005, fuechsle_2009, virtanen_2010, bergeret_2011, dassonneville_2018, basset_2019, dubos_2001a, haxell_2022}. The adiabatic changes in the ABS energies, as well as the multiple transitions induced by microwave photons between the ABSs or between the ABSs and the continuum, can result in a non-trivial dynamics of the supercurrent-carrying states.~\cite{zaikin_1983, cuevas_2002, bergeret_2011, virtanen_2010} Such effects can give rise to highly distorted CPRs, which exhibit sign-reversals of the supercurrent and $\pi$-periodic oscillations at twice the Josephson frequency.~\cite{argaman_1999, kroemer_1999, lehnert_1999, dubos_2001a, baselmans_2002, jacobs_2005, dou_2021}. 

We notice that in the experiment the induced gap $\Delta^* \simeq \SI{280}{\micro\eV}$ (\SI{67}{\GHz}), so that we cover values $hf \simeq 0.03~\Delta^*$. In an effort to capture the emergence of half-integer Shapiro steps, we describe the junction dynamics by adiabatically incorporating non-equilibrium effects into the RCSJ model of Fig.~\ref{fig1}c through a single effective CPR. The latter is provided by the thick grey line in Fig.~\ref{fig3}e, and its origin will be discussed later. In Figure~\ref{fig3}d, we plot the step width for integer and half-integer values of $n$ vs. $P_{RF}$, obtained as horizontal slices of Figure~\ref{fig2}c at constant $V$. The dashed black line in the figure shows the results of the simulation using the effective non-equilibrium CPR, while the dashed grey line represents the equilibrium one. Although the equilibrium CPR effectively reproduces the integer steps in the oscillatory pattern (bottom row), it completely fails to capture the half-integer steps (top row). This is despite the presence of higher-order harmonics in the highly skewed CPR, which are often attributed to the origin of half-integer steps~\cite{ueda_2020}.

To gain further insight into the origin of such a distorted CPR, we used a tight-binding method within the Keldysh-Green's function approach~\cite{cuevas_2002, bergeret_2011} to numerically calculate the current-phase relationship of an SNS junction irradiated by a microwave tone. The model describes a single-channel Josephson junction with an arbitrary junction transparency $\tau$ and gap $\Delta^*$. The microwave driving is included as a time-dependent modulation of the phase difference across the junction with amplitude $w = e V_{RF}/hf$. Figure~\ref{fig3}e shows the simulated CPR for microwave irradiation of $hf = 0.1~\Delta^*$ at a microwave driving of $w \sim 1.3$ and $\tau = 0.98$ for different temperatures. The dashed grey line represents the equilibrium CPR. The microwave irradiation significantly alters the CPR, boosting a strong second harmonic, which results in the development of an additional minimum. This provides insight into the origin of the effective non-equilibrium CPR used in the RCSJ model. The wiggles in the CPR are due to non-equilibrium population of Floquet sidebands produced by the microwave driving and disappear at temperatures on the order of the driving frequency, $k_BT\sim hf$. In turn, the secondary minimum is robust and still visible at $k_B T = 0.1~\Delta^*$, as shown in Figure~\ref{fig3}e, and it qualitatively agrees with the robustness of the half-integer steps with respect to temperature. In the Supporting Information, we detail the theoretical model and present additional simulations showing that reducing the junction transparency results in the disappearance of the CPR's secondary minimum (Figure~S10).

The outlined procedure should be regarded as an attempt to reconcile the results of the adiabatic approximation, typical of the RCSJ model, with the microscopically calculated CPR in the presence of microwave driving and in the absence of a steady voltage across the junction. In particular, the model reproduces the two-minima shape of the effective non-equilibrium CPR only within a limited range of $w$ values, which is inconsistent with the experimental observations and highlights the limitations of the present description. Alternative phenomenological theories of non-equilibrium supercurrents have been proposed~\cite{lehnert_1999, argaman_1999, kroemer_1999}, which model the system by considering both the ABSs and their occupation distributions oscillating at the Josephson frequency. The specific structure of the ABSs, including the effects of finite junction length or ballistic quasi two-dimensional transport, may be responsible for the discrepancies between different predictions, which calls for more comprehensive theories.

In conclusion, we have investigated a highly transmissive Josephson junction made of an InSb nanoflag with Nb contacts. Our results indicate strong evidence of parallel transport in both long and short conducting channels, confirmed by the temperature-dependent supercurrent and magnetic field interference. Under microwave irradiation, we observe strong half-integer Shapiro steps, showing a non-monotonic temperature evolution that points to non-equilibrium effects induced by the driving. The observed phenomenology is only partially captured by the predictions based on the adiabatic approximation in terms of a non-equilibrium CPR. Further theoretical developments are needed to address the presence of strong second harmonic supercurrents in ballistic, highly transparent SNS junctions. Future experiments should investigate the potential of InSb nanoflag Josephson junctions for exploring the coherent manipulation of Andreev states and their non-equilibrium dynamics.

\begin{acknowledgement}
The authors thank Daniele Ercolani for his help with the growth of the InSb nanoflags and Michal Nowak for useful discussions. This research activity was partially supported by the FET-OPEN project AndQC (H2020 Grant No. 828948). E.S. and F.G. acknowledge the EU’s Horizon 2020 research and innovation program under Grant Agreement No. 800923 (SUPERTED) and No. 964398 (SUPERGATE) for partial financial support.
\end{acknowledgement}


\begin{suppinfo}
\setcounter{figure}{0}
\setcounter{equation}{0}
\setcounter{section}{0}
\renewcommand\thesection{S\arabic{section}}
\renewcommand\thefigure{S\arabic{figure}}
\renewcommand\theequation{S\arabic{equation}}

\section{Sample Information and Measurement Techniques}
The InSb nanoflags utilized in this work have been extensively described in previous studies~\cite{verma_2020, verma_2021}. They are defect-free structures that exhibit excellent electrical properties, including high mobility (up to \SI{29500}{\cm\square\per\V\per\second}) and a large mean free path ($l_e \simeq \SI{500}{\nm}$) at $T = 4.2$~K. The devices were fabricated by placing nanoflags on a $p$-doped Si/SiO$_2$ substrate, which serves as a backgate, and connecting them with \SI{150}{\nm} of Nb. A passivation step was performed prior to metal deposition to improve the semiconductor-metal transparency. Further information on device fabrication can be found in the supporting material of~\citet{salimian_2021, turini_2022}. We conducted transport measurements using a low-temperature Leiden Cryogenics dilution refrigerator with a base temperature of \SI{75}{\milli\K}. The cryostat is equipped with a three-level filtering system, comprising $\pi$ filters at room temperature as well as cryogenic $\pi$ and $RC$ filters at base temperature. The $V(I)$ curves were acquired in a standard four-wire configuration, with the junction current-biased using a Yokogawa GS200 voltage source over a \SI{10}{\Mohm} resistor. The voltage drop over the junction was amplified by a factor $1000$ using a room temperature DL1201 voltage preamplifier operated in battery mode and acquired by an Agilent 34410 multimeter. We applied voltage to the backgate using a Keithley 2602 voltage source. Microwave signals were applied using an R\&S SMR20 microwave source to an attenuated semi-rigid open-ended coaxial cable in close proximity to the sample holder, which had been attenuated by 20 dB and 10 dB at 3K and Cold plates, respectively. For measurements in magnetic field, we used a low-noise Keithley 2400 sourcemeter connected to a \SI{2}{\tesla} American Magnetics magnet.

\section{Dependence on Backgate Voltage}

Figure~\ref{fig_si_backgate_1}(a-d) displays $V(I)$ curves in a higher current range at different backgate voltages $V_G$ ranging from \SI{10}{\V} to \SI{40}{\V}. The black dashed line fitted to the Ohmic region allows obtaining the normal state resistance $R_N$ and excess current $I_{exc}$. In panel (e), we present the switching current $I_{sw}$ and retrapping current $I_{rt}$ over the same range of gate voltages. The switching current $I_{sw}$ decreases from \SI{170}{\nA} to \SI{75}{\nA} as the gate voltage is varied, while the retrapping current $I_{rt}$ remains roughly constant at approximately \SI{30}{\nA}.

\begin{figure}[H]
\centering
\includegraphics{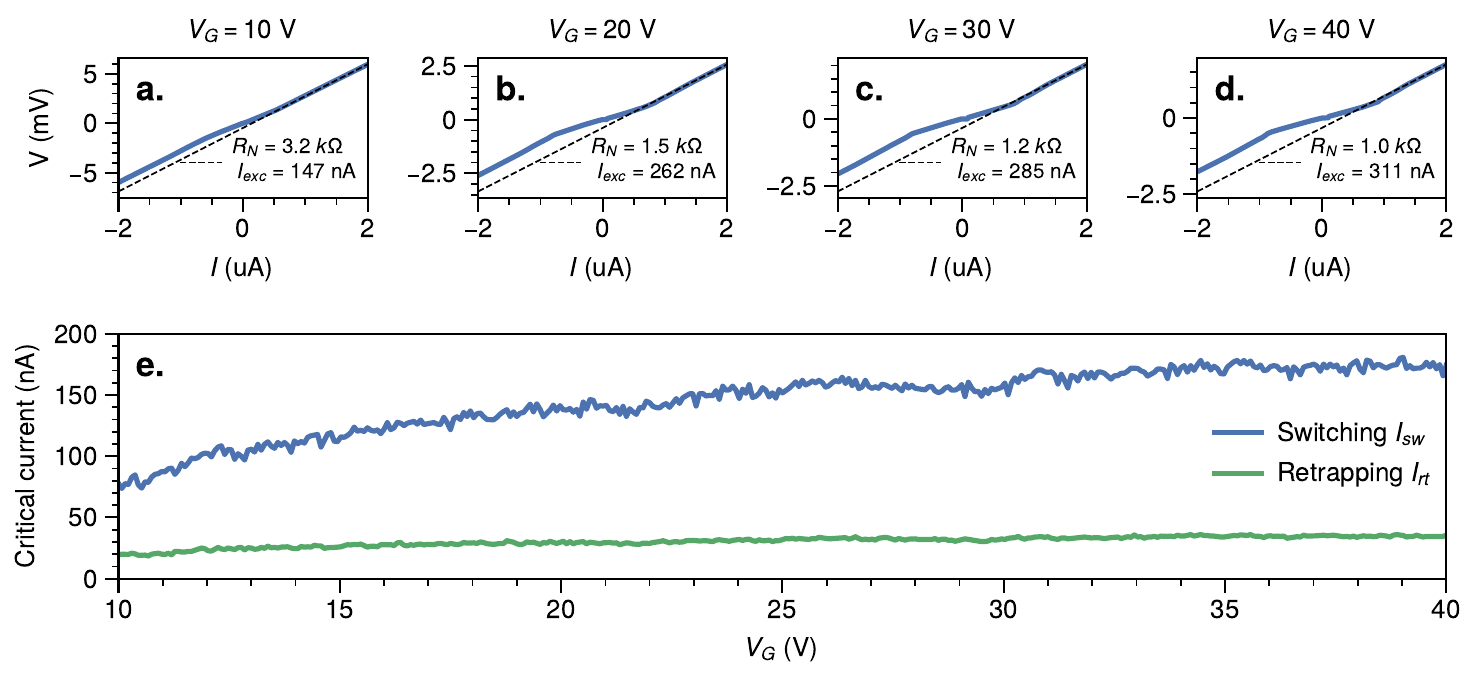}
\caption{(a-d) Sample I-V traces for different $V_G$ values (10,20,30,40) \si{\V}. The black dashed line is the linear fit to the Ohmic region used to extract $R_N$ and $I_{exc}$. (e) Full backgate dependence of the switching and retrapping current in the same gate voltage range.}
\label{fig_si_backgate_1}
\end{figure}

Figure~\ref{fig_si_backgate_2}a depicts the evolution of $R_N$ and $G_N = 1/R_N$ as a function of $V_G$, with the resistance decreasing from \SI{3}{\kohm} (at $V_G = \SI{10}{\V}$) to \SI{1}{\kohm} (at $V_G = \SI{40}{\V}$). Figure~\ref{fig_si_backgate_2}b shows the excess current as a function of $V_G$ in the same range. The product $I_{exc}R_N$ remains roughly constant at about \SI{350}{\uV} ($1.2~\Delta^*/e$) over the entire gate voltage range, close to the theoretical value of $8/3~\Delta^*/e$ predicted for the ballistic case (Fig.~\ref{fig_si_backgate_2}c)~\cite{flensberg_1988}. Moreover, the product $I_{sw} R_N$ is expected to be $I_{sw} R_N = 10.82~E_{Th}/e$ for a long diffusive junction in the limit $\Delta^* \gg E_{Th}$~\cite{dubos_2001}. In our device, $I_{sw} R_N$ varies from \SI{180}{\uV} to \SI{250}{\uV} ($7-10~E_{Th}/e$) (Fig.~\ref{fig_si_backgate_2}d).

\begin{figure}[H]
\centering
\includegraphics{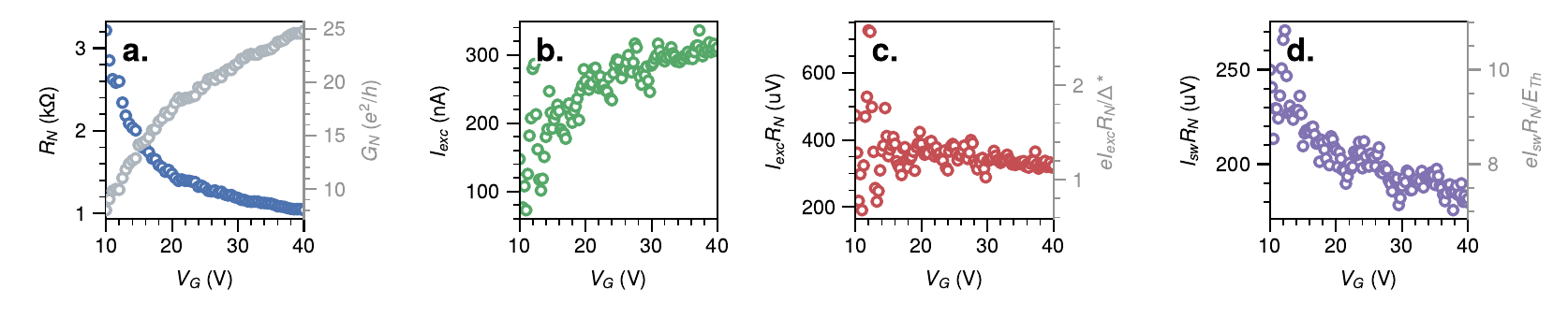}
\caption{(a) Normal state resistance $R_N$, (b) excess current $I_{exc}$, (c) $I_{exc}R_N$ product, and (d) $I_{sw} R_N$ product plotted as functions of the backgate voltage $V_G$.}
\label{fig_si_backgate_2}
\end{figure}

\section{Magnetic Interference Maps}

Figures~\ref{fig_si_fraunhofer_vs_T} and \ref{fig_si_fraunhofer_vs_Vg} present additional measurements of magnetic interference patterns at different temperatures, $T = (300, 500, 800)~\si{\milli\K}$, and backgate voltages, $V_G = (30, 20, 10)~\si{\V}$. The magnetoresistance maps in Figure~\ref{fig_si_fraunhofer_vs_T} demonstrate a decrease in the Gaussian-like contribution as the temperature increases, further confirming that the exponential suppression of $I_{sw}$ is related to states in the long junction limit. The Fraunhofer diffraction side-lobes remain unchanged up to a temperature of \SI{500}{\milli\K}, consistent with the limited dependence of $I_{sw} (T)$ for modes in the short junction limit, and eventually begin to disappear only for $T > \SI{800}{\milli\K}$. 

\begin{figure}[H]
\centering
\includegraphics{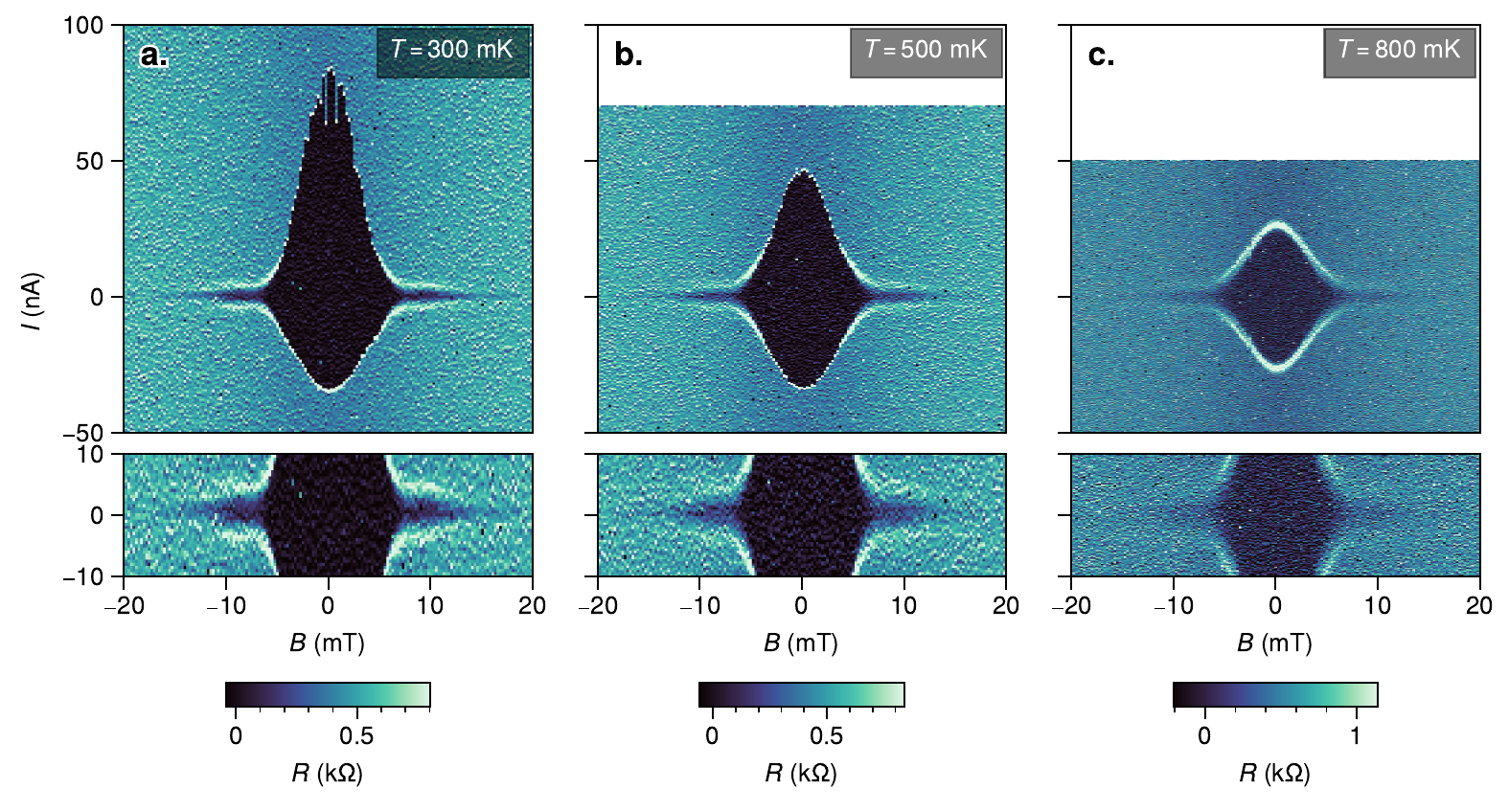}
\caption{Magnetoresistance maps at temperatures of (a) \SI{300}{\milli\K}, (b) \SI{500}{\milli\K}, and (c) \SI{800}{\milli\K}. Below each panel, a zoom-in on the low bias region shows the behavior of the side-lobes with respect to $T$.}
\label{fig_si_fraunhofer_vs_T}
\end{figure}

Figure~\ref{fig_si_fraunhofer_vs_Vg} shows the magnetoresistance evolution in $V_G$. States in the long junction limit are more sensitive to changes in the semiconductor depletion level at low gate voltages $V_G$, as evidenced by the corresponding changes in the Gaussian amplitude. Only a little variation of the Fraunhofer diffraction lobes is observed confirming the high homogeneity of the current density in the short junction area between the electrodes. 

\begin{figure}[H]
\centering
\includegraphics{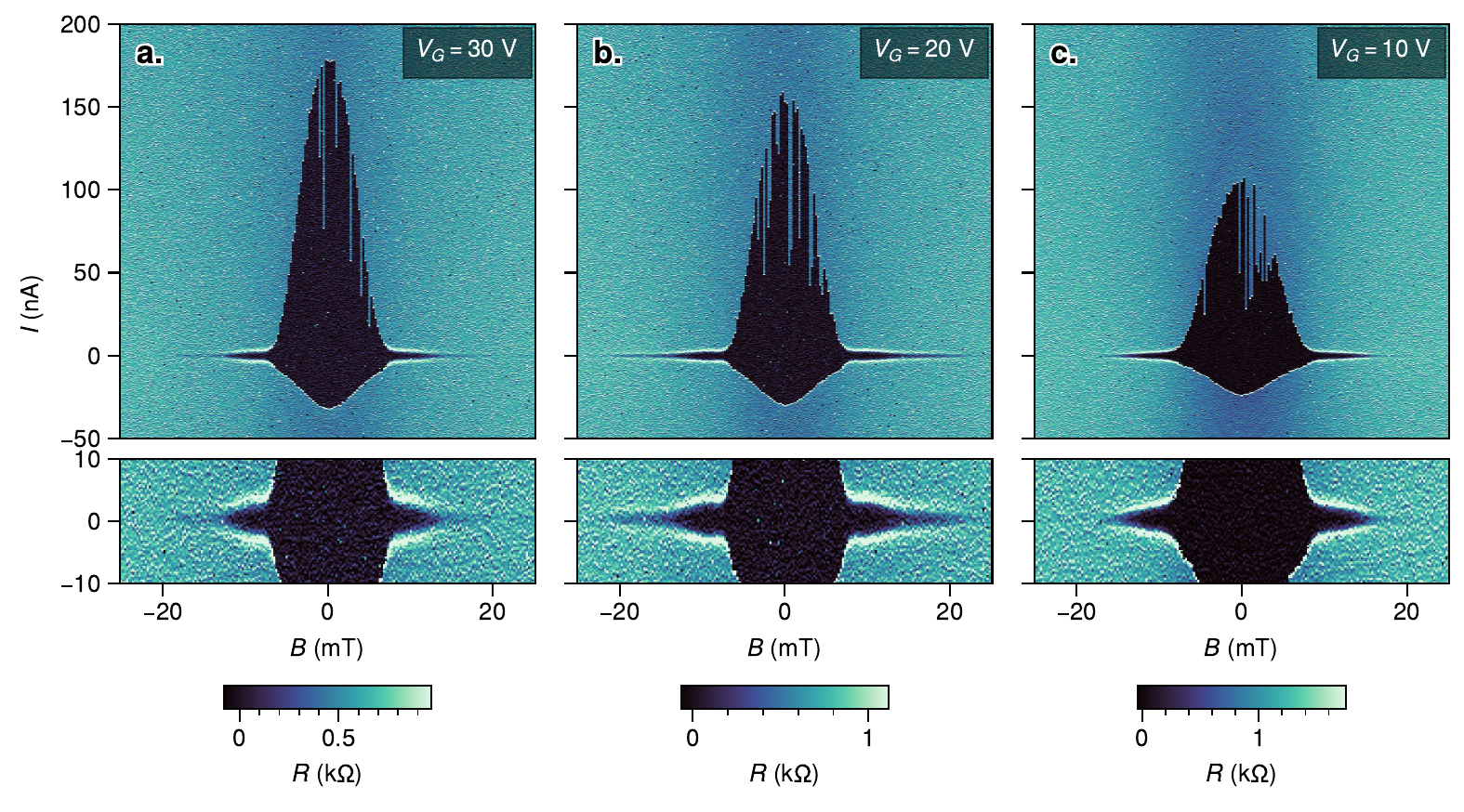}
\caption{Magnetoresistance maps at backgate voltage $V_G$ of (a) \SI{30}{\V}, (b) \SI{20}{\V}, and (c) \SI{10}{V}. Below each panel, a zoom-in on the low bias region shows the behavior of the side-lobes with respect to $V_G$.}
\label{fig_si_fraunhofer_vs_Vg}
\end{figure}

\section{Shapiro Maps at finite Magnetic Field}
Figure~\ref{fig_si_shapiro_B} depicts the Shapiro maps obtained at a frequency of $f = \SI{1.75}{\GHz}$ for different out-of-plane magnetic field values ranging from 0 to 11~\si{\milli\tesla}.  As the magnetic field suppresses $I_c$, the reduced drive frequency $\Omega = \frac{2\pi f}{2e I_c R_j/\hbar}$ increases, and the $V(I)$ maps follow the Bessel function dependence on the applied RF power~\cite{russer_1972, larson_2020}.
\begin{figure}[H]
\centering
\includegraphics{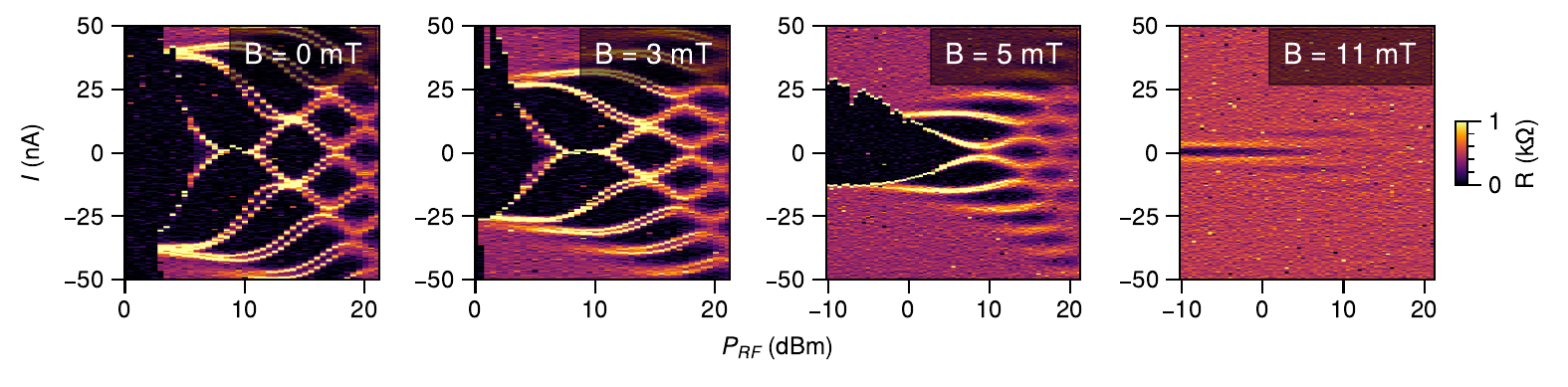}
\caption{Shapiro maps at $f = \SI{1.75}{\GHz}$ for various out-of-plane magnetic field strengths $B$ ranging from 0 to 11~\si{\milli\tesla}.}
\label{fig_si_shapiro_B}
\end{figure}

\section{Shapiro Maps at different Backgate Voltages and Temperatures}
The evolution of the half-integer steps is robust in temperature and backgate voltage, as detailed by the scans in Fig.~\ref{fig_si_shapiro_Vg_and_T} for $V_G = \SI{40}{\volt}$ (top row) and $V_G = \SI{10}{\volt}$ (bottom row).
\begin{figure}[H]
\centering
\includegraphics{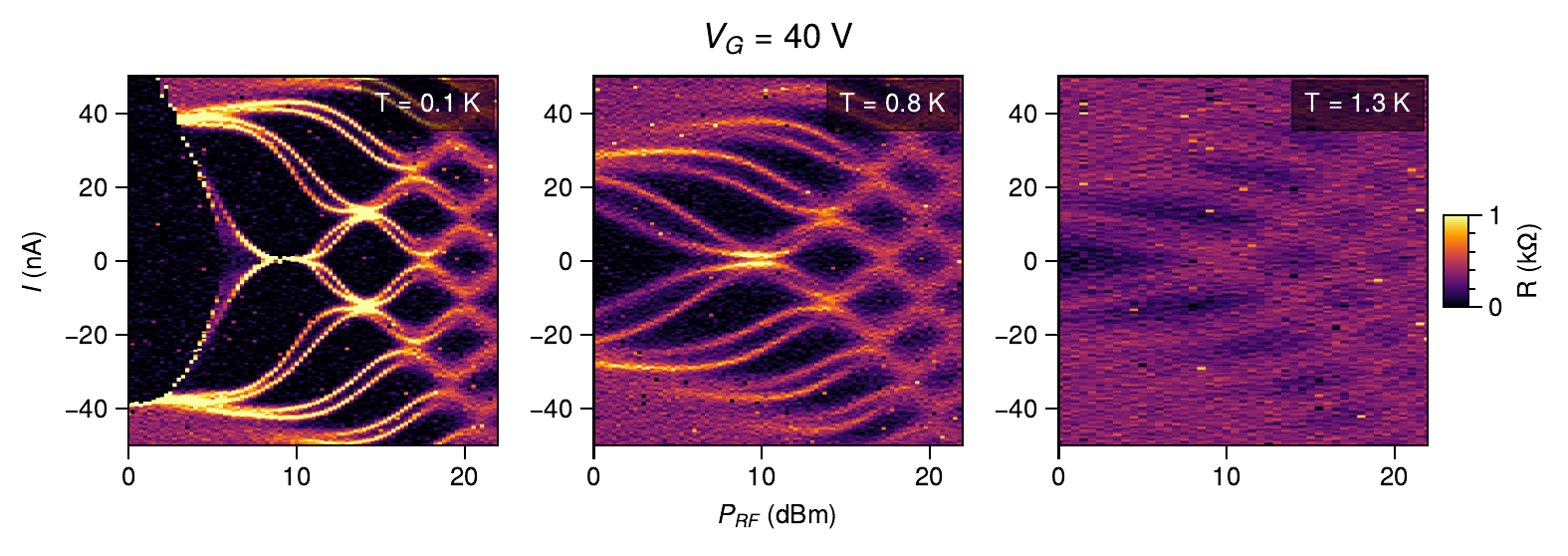}
\includegraphics{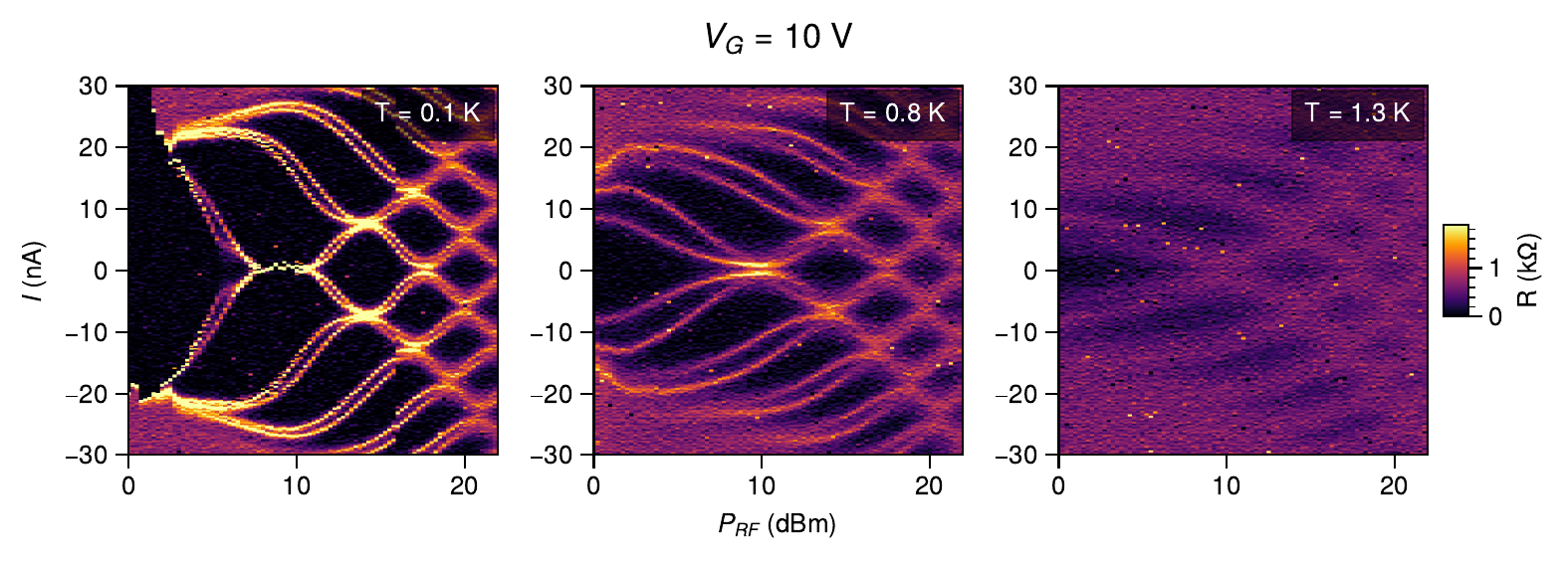}
\caption{Top row: Shapiro maps at $f = \SI{1.75}{\GHz}$ for different temperatures of \SI{0.1}{\K}, \SI{0.8}{\K}, and \SI{1.3}{\K} at a backgate voltage of \SI{40}{\V}. Bottom row: same maps for a backgate voltage of \SI{10}{\V}.}
\label{fig_si_shapiro_Vg_and_T}
\end{figure}

\section{Shapiro Maps at Zero Crossing Step}

We present in Fig.~\ref{fig_si_shapiro_zero_crossing} a more detailed scan of the Shapiro map at $f = \SI{1.75}{\GHz}$, covering a restricted range of microwave power and currents as shown in Figure~2 of the main text. Zero-crossing steps (red line) are visible as a result of the overlapping $\pm 1$ lobes. The presence of zero-crossing steps has been extensively investigated in~\citet{larson_2020} and explained as a consequence of the shunting $RC$ environment and the high $I_c R_j$ product.

\begin{figure}[H]
\centering
\includegraphics{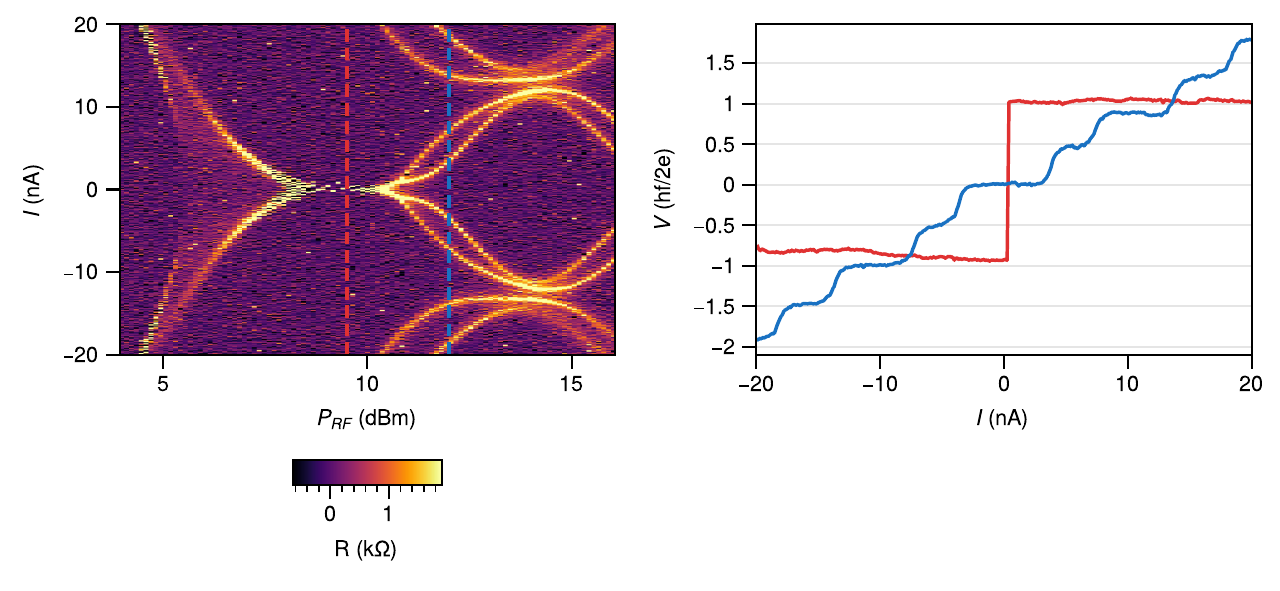}
\caption{Left: A zoom-in view of the Shapiro maps at $f = \SI{1.75}{\GHz}$, as shown in Figure~2 of the main text, highlighting the zero-crossing steps and providing finer details of the half-integer plateaus. Right: The red and blue $V(I)$ cut shown on the left for $P_{RF} = 9~\si{dBm}$ and $P_{RF} = 12~\si{dBm}$, respectively.}
\label{fig_si_shapiro_zero_crossing}
\end{figure}

\section{Data from an additional Device with lower Transparency}
An additional device has been measured in a similar way, which had a \SI{5}{\nm} Ti adhesion layer embedded under the niobium layer. In this case, only weak signatures of half-integer steps are visible.

\begin{figure}[H]
\centering
\includegraphics{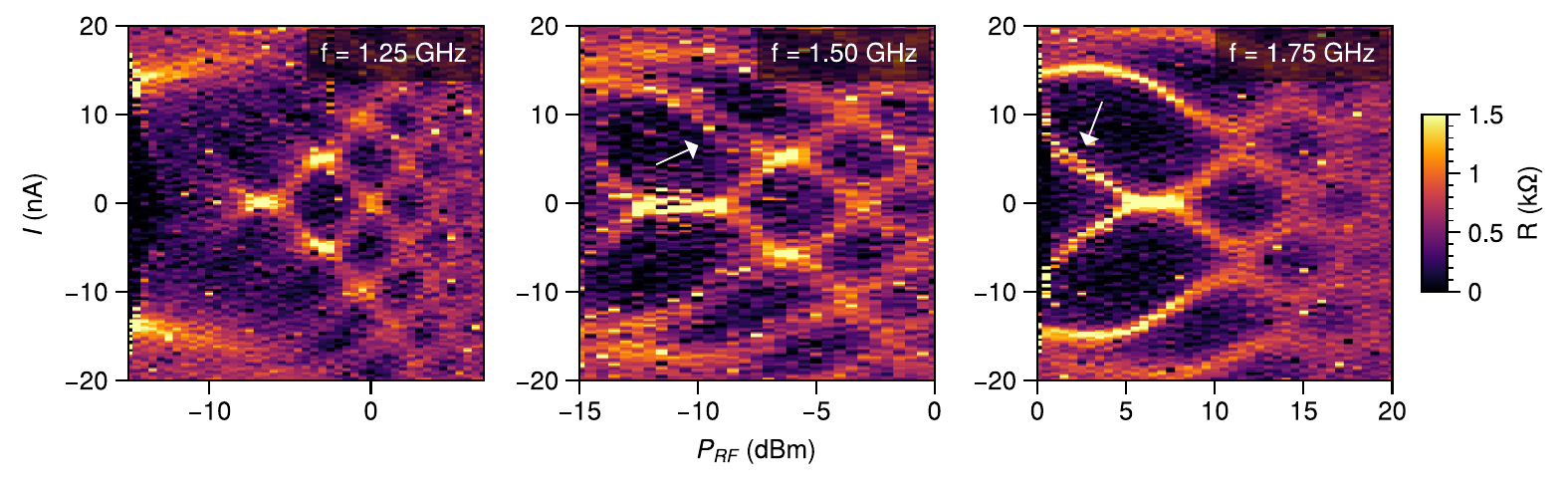}
\caption{Shapiro maps for $f = \num{1.25}, \num{1.50}$ and \SI{1.75}{\GHz} for a device with reduced transparency.}
\label{fig_si_shapiro_additiona_device}
\end{figure}

\section{Theory} \label{sec:num_sim}
\subsection{CPR under microwave irradiation}
Our microscopic model for a microwave-irradiated Josephson junction is based on the works of~\citet{cuevas_2002, cuevas_2006, bergeret_2011} and consists of a highly transparent junction, with a short ballistic region between the left and right superconducting leads (L and R, respectively), as schematically displayed in Figure~\ref{fig_si_noneq_model}a.
\begin{figure}
	\centering
 \includegraphics{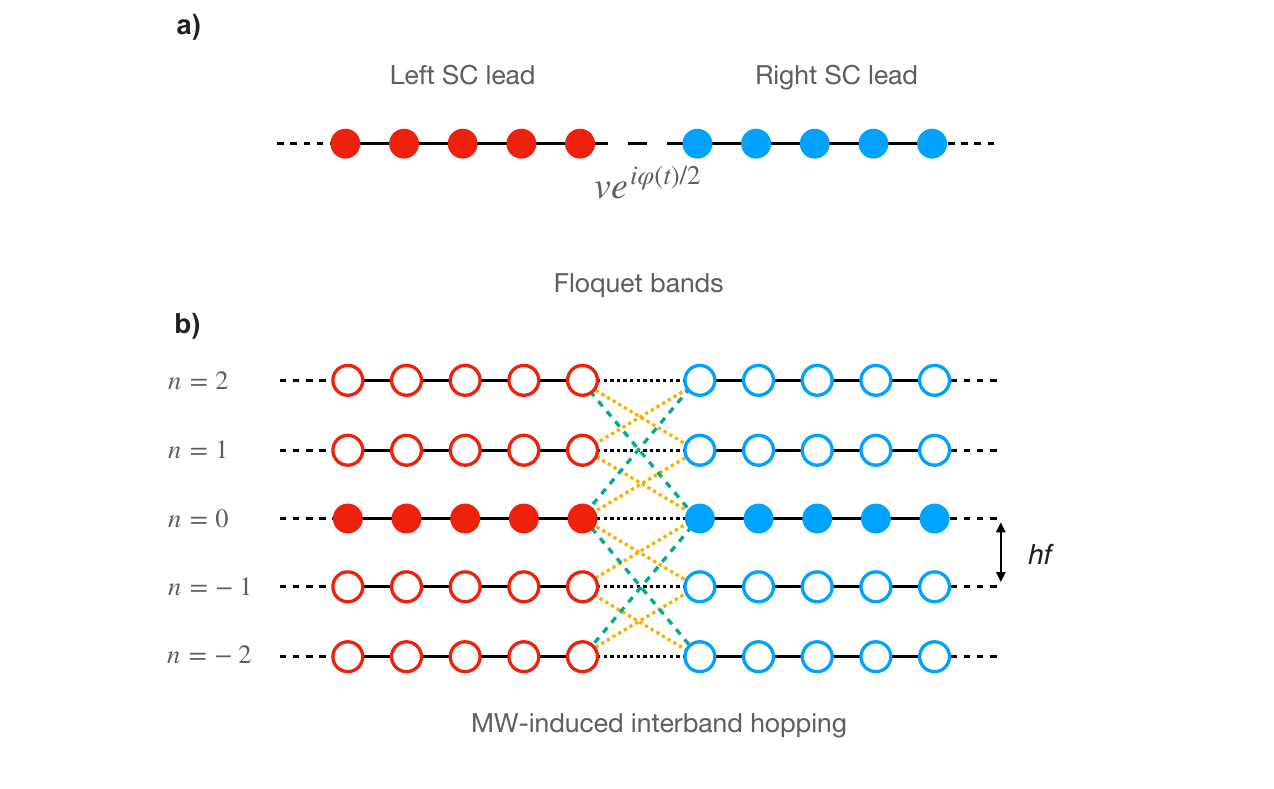}
	\caption{(a) 1D tight-binding model describing the Josephson junction. The driving appears as a time-dependent phase in the left-right hopping amplitude. (b) Floquet side bands shifted in energy by $nhf$. The different bands are coupled at the interface by the time-oscillating phase.}
	\label{fig_si_noneq_model}
\end{figure}
We describe the system through the 1D tight-binding Hamiltonian $H=H_L+H_R+\sum_\sigma (v c^\dag_{L\sigma}c_{R\sigma}+v^* c^\dag_{R\sigma}c_{L\sigma}),$ where the left and right leads are described by $H_\alpha=-\mu\sum_{n\sigma}c^\dag_{\alpha,n,\sigma}c_{\alpha,n,\sigma}-t\sum_{n,\sigma}(c^\dag_{\alpha,\sigma,n}c_{\alpha,\sigma,n+1}+{\rm H.c.})+\Delta_\alpha\sum_nc^\dag_{\alpha,n,\uparrow}c^\dag_{\alpha,n,\downarrow}+{\rm H.c.}$, with $c$ and $c^\dag$ being the annihilation/creation operators for particles with spin $\sigma$ in the superconducting leads and $\Delta_\alpha=\Delta e^{i\varphi_\alpha}$, with $\varphi_R-\varphi_L=\varphi$ their phase difference. The normal state transmission of this single channel model is ${\tau}=\frac{4(v/W)^2}{(1+(v/W)^2)^2}$, where $W=1/(\pi\rho_F)=\sqrt{4t^2-\mu^2}/2$ and $\rho_F$ is the density of states at the Fermi energy in the leads. The current takes the form $\hat{I}(t)=\frac{ie}{\hbar}\sum_\sigma (v c^\dag_{L\sigma}c_{R\sigma}-v^* c^\dag_{R\sigma}c_{L\sigma})$, and under microwave irradiation, the phase difference acquires the time dependence $\varphi(t)=\varphi_0+2w\sin(2\pi f t)$ with $w=eV_{\rm ac}/hf$, so that $v\to v e^{i\varphi(t)/2}$.

The microwave source can inject and absorb photons of frequency $f$, so that an incident carrier with energy $\epsilon$ can be scattered into states with energy $\epsilon+nhf$. Introducing the Floquet sidebands, which are replicas of the system shifted in energy by $nhf$, the hopping term $\hat{v}_{LR}$ can couple different sidebands $\hat{v}_{n,m}=\int dt e^{i(n-m)2\pi f t}\hat{v}(t)=v \begin{psmallmatrix}
J_{n-m}(w)e^{i\varphi_0/2} & 0\\
0 & -J_{m-n}(w)e^{-i\varphi_0/2}
\end{psmallmatrix}$, where $J_n(w)$ are Bessel functions of the first kind, and we have absorbed the possible phase of $v$ in the phase difference $\varphi_0$. The hopping between the L and R leads acquires a matrix structure that connects the rightmost site in the L lead and Floquet band $n$ with the leftmost site of the R lead and Floquet band $m$, as schematized in Fig.~\ref{fig_si_noneq_model}b.

\begin{figure}
    \centering
\includegraphics{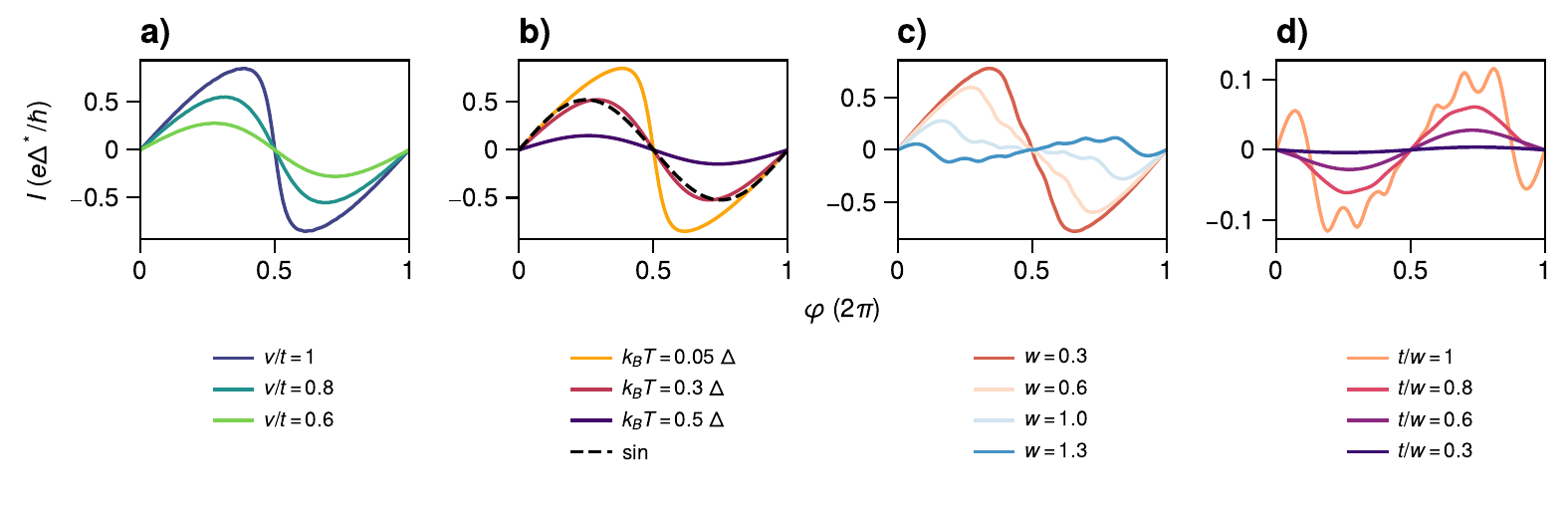}
    \caption{Different CPRs: (a) in the absence of microwave irradiation for different transparencies, (b) in the absence of microwave irradiation for different temperatures, (c) for different microwave drive strengths at $v/t=1$, and (d) for different transparencies at driving $w=1.3$.}
    \label{fig_si_noneq_theory}
\end{figure}

We now apply the microscopic theory and calculate different CPRs under microwave irradiation. In the absence of driving ($w = 0$), the model accurately reproduced the CPRs of the junction in both the highly transparent and tunneling regimes by varying the ratio $v/t$, as shown in Fig.~\ref{fig_si_noneq_theory}a, and for different temperatures, as shown in Fig.~\ref{fig_si_noneq_theory}b. We then apply increasing microwave driving and observe a second zero of the CPR in the interval 0-$\pi$, along with a region of negative current for positive phase bias, which indicated the occurrence of a secondary minimum (Fig.~\ref{fig_si_noneq_theory}c). At high driving amplitude, the CPR exhibits wiggles due to the presence of Floquet sidebands, which decay with temperature, as shown in Fig.~3e of the main text. Finally, Figure~\ref{fig_si_noneq_theory}d shows that the CPR loses its secondary zero for low transmission. These features are qualitatively similar to those observed in the additional device reported in the experiment. Including additional bands up to 6 does not qualitatively change the discussion above.

\subsection{RCSJ model}
In the previous section, we described the microscopic theory of non-equilibrium supercurrents in a microwave irradiated Josephson junction in the presence of an AC voltage bias. However, modeling the time-dependent phase dynamics in the presence of non-equilibrium effects, particularly for the experimentally relevant current-bias scenario, is more challenging. Moreover, we find that the environment surrounding the junction plays an important role, as confirmed by the deviations from the Bessel regime and the presence of zero-crossing steps. We opted for a simplistic approach that employs a modified version of the resistively and capacitively shunted junction (RCSJ) model, which includes the dissipative environment surrounding the junction~\cite{jarillo-herrero_2006, larson_2020} and incorporates the non-equilibrium effects only in a single effective CPR. Despite the simplicity of the assumptions, we are able to capture the main findings of this work. The junction, which has a critical current $I_c$, is shunted by a capacitance $C_j$ and resistance $R_j$, and is additionally shunted by an RC environment represented by a parallel capacitor $C$ and resistor $R$. The entire circuit is biased by a current $I$, which accounts for the external DC and AC bias. The equations for the current $I$ and the voltage $V$, shown in Figure 1c, are:
\begin{equation}
    \begin{aligned}
        I&=I_{DC}+I_{RF} \sin{(2 \pi f t)}\\
        &=C \frac{dV}{dt}+  \text{CPR}(\varphi) +\frac{\hbar}{2 e R_j}\frac{d\varphi}{dt}+\frac{\hbar C_j}{2 e}\frac{d^2 \varphi}{dt^2}\\
        V&=\frac{\hbar}{2e}\frac{d\varphi}{dt}+R\left(\text{CPR}(\varphi) +\frac{\hbar}{2 e R_j}\frac{d\varphi}{dt}+\frac{\hbar C_j}{2 e}\frac{d^2 \varphi}{dt^2}\right),
    \end{aligned}
    \label{eq_RCSJ}
\end{equation}
where $\varphi$ is the macroscopic phase difference across the junction, $I_{DC}$ and $I_{RF}$ are the DC and RF current biases, respectively, and $\text{CPR}(\varphi)$ is the junction's current-phase relationship. We use a fourth-order Runge-Kutta method to solve for $\varphi(t)$ and obtain the DC voltage across the junction as $V_j=\left<\frac{\hbar}{2e}\frac{d\varphi}{dt}\right>$. 

In the limit of small $I_c$, the Shapiro map follows the Bessel function dependence, with steps centered at $I_n = \frac{V_n}{R_j}$ and an extension of $\sim I_c |J_n(2w)|$, where $V_n = n \frac{hf}{2e}$ and $w=eV_{AC}/hf$. Figure~\ref{fig_si_rcsj}a shows the Shapiro map at lower $I_c$ with an applied external magnetic field of $B = \SI{5}{\milli\tesla}$. From the position of the centers, we can extract $R_j \sim \SI{420}{\ohm}$ (dotted white lines), while the dashed lines show a good agreement with the Bessel behavior, depicting the amplitudes $I_c |J_n(2w)|$, with $V_{AC} = \alpha \times 10^{P_{RF}/20}$, $\alpha \sim 0.8$ and $I_c \sim \SI{10}{\nA}$.

Figures~\ref{fig_si_rcsj}b and \ref{fig_si_rcsj}c show the complete maps for the simulation presented in Fig.~3d of the main text for the equilibrium and non-equilibrium CPRs, respectively, obtained by the model in Eq.~\ref{eq_RCSJ}. While the equilibrium CPR well describes the overall trend, it completely lacks half-integer steps, which are instead captured by the effective non-equilibrium CPR. This is despite the presence of higher-order harmonics in the skewed equilibrium CPR. In the simulation, we estimate the geometric capacitance of the junction to be $\sim \si{\fF}$ and neglect $C_j$. The capacitance $C$ is determined by the bonding pads' capacitance to the SiO$_2$ backgate, which we estimate to be $C \sim \SI{15}{\pF}$, while the value of $R$ is set to $R \sim \SI{150}{\ohm}$ to achieve the best agreement with the experiment. The CPR is expressed as $\text{CPR}(\varphi) = \sum_n I_{c,n} \sin(n\varphi)$, where $I_c = \max_\varphi \text{CPR}(\varphi)$, which is set to \SI{35}{\nA}. The current $I_{RF}$ is given by $I_{RF} = \beta \times 10^{P_{RF}/20}$ with $\beta \sim 20$.
\begin{figure}[H]
\centering
\includegraphics{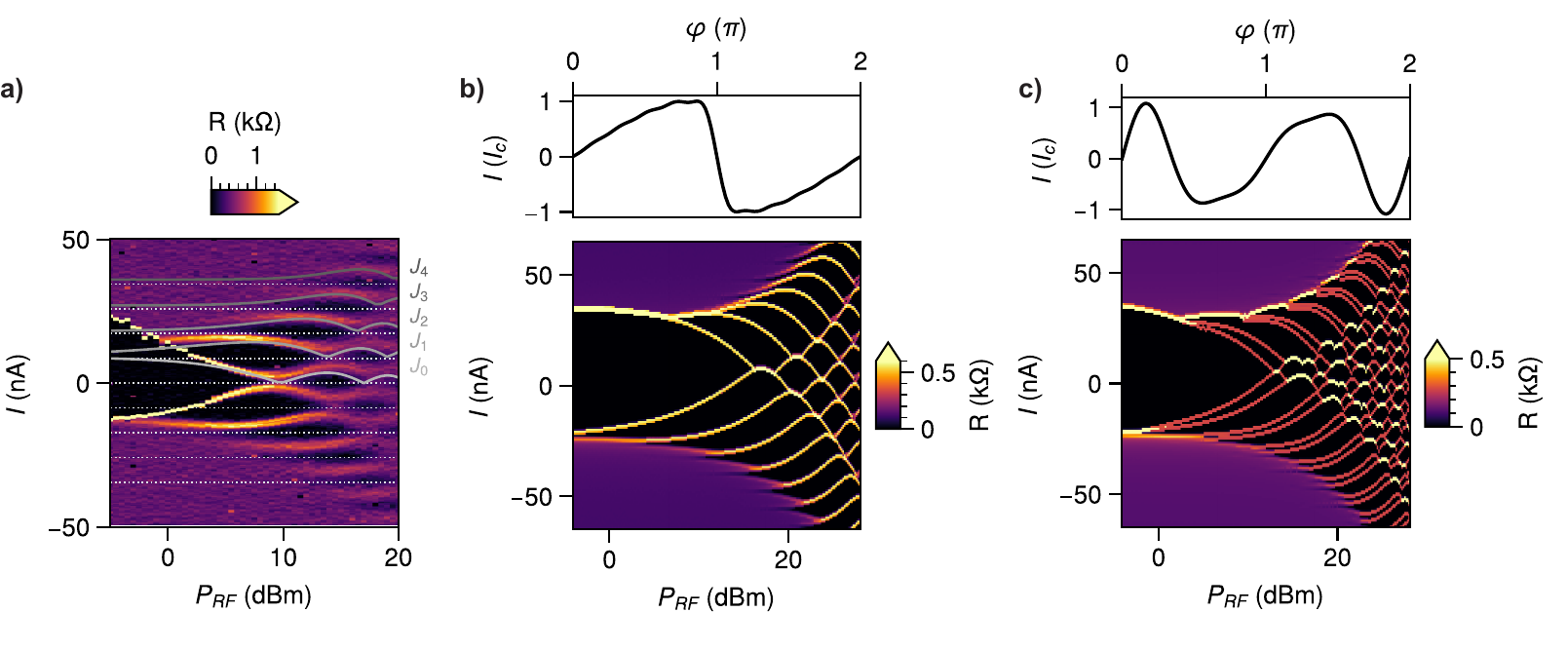}
\caption{(a) Experimentally measured Shapiro map in the Bessel regime at $B = \SI{5}{\milli\tesla}$. The dotted white lines are used to extract $R_j$, while the continuous lines show the dependence on the Bessel function $J_n$. (b) Numerical simulation of the Shapiro map with the equilibrium skewed CPR. The amplitudes of the harmonics are $I_{c,n} =$ (0.91, -0.33, 0.19, -0.12, 0.09, -0.06, 0.04, -0.03, 0.03, -0.02). (c) Same as in (b) for the effective non-equilibrium CPR with the amplitudes $I_{c,n}$ = (-0.38, 0.75, 0.44, 0.22).}
\label{fig_si_rcsj}
\end{figure}
\end{suppinfo}

\bibliography{bibliography.bib}
\end{document}